\documentclass[12pt,a4paper]{article} 	%tipo documento
\usepackage[T1]{fontenc}		     	%caratteri accentati e particolari
\usepackage[english]{babel}
\usepackage{graphicx}			    	%inserire immagini
\usepackage{indentfirst}		    	        %indenta primo paragrafo
\usepackage{booktabs}				%tabelle
\usepackage{amsmath}			%roba matematica
\usepackage{siunitx}				%unità sistema internazionale
\usepackage{mathrsfs}              	       %lettere strane matematica
\usepackage{euscript}
\usepackage{geometry}          	       %serve per i margini altrimenti manualmente \noindent
\usepackage{amsmath}
\usepackage{float}

\title{\huge \textbf{Photometry of Delta Scorpii from 1996 to 2013 using SOHO LASCO C3 coronograph}}

\author{\small{Costantino Sigismondi 1,2,3, Graziano Ucci, Vanessa Zema, Francesco Scardino, Federico Maria Vincentelli 4}\\
\small{\textit{1 ICRANet, International Center for Relativistic Astrophysics Network, Rome (Italy)}}\\ 
\small{\textit{2 Galileo Ferraris Institute and Pontifical Athenaeum Regina Apostolorum, Rome (Italy)}}\\ 
\small{\textit{3 Observatorio Nacional and Universidade Federal do Rio de Janeiro (Brazil)}}\\
\small{\textit{4 Department of Physics, Sapienza University of Rome, (Italy)}}}

\begin{document}

\maketitle

\begin{abstract}

The variabile star Delta Scorpii is in conjunction with the Sun at the end of November each year. We studied its magnitude by averaging the observations of 28 Nov - 1 Dec from 1996 to 2013 using the coronograph LASCO C3 on-board the SOHO Satellite and we extended of four years, i.e. 25\% of the total light curve, back to 1996, with respect to the present AAVSO dataset on this star. 
The 0.2 magnitude scatters of the single measurements have been studied and the sources of such disturbances are vignetting and diffraction patterns from the coronograph.
The new data collected on Delta Scorpii show its minimum at mv=2.5 magnitudes for 1996 and 1997, confirming the values observed during the minimum of 2009, and the main periodicity of 11 years in the stellar variability.

\end{abstract}

\section{Introduction}
Delta Scorpii is a second magnitude Be giant; it is a double star and the atmospheres are grazing at periastron. Delta Scorpii shows an irregular and eruptive variability with 11 years of main periodicity linked to the orbital period\cite{SIGISMONDI}. The variability of Delta Scorpii has been studied since the year 2000\cite{OTERO}. In this paper we examine an homogeneus dataset from SOHO satellite to extend the study of this star back to 1996.
Despite its main purpose of observing the Sun, SOHO (SOlar Heliospheric Observatory) satellite is here used to perform differential photometry in its 17$^{\circ}$ field of view. In the last days of November and beginning of December the Sun approaches Delta and Alpha Scorpii, and they appear in the field of view of the SOHO LASCO C3 coronograph.

\section{The C3 Coronograph of SOHO}
The Large Angle and Spectrometric Coronagraph LASCO is a set of three coronagraphs (C1, C2 and C3) on-board the SOHO Satellite.

	\begin{figure}[H]
 	   	 	 	\centering
 	   	 	 	\includegraphics[width=0.95\linewidth]{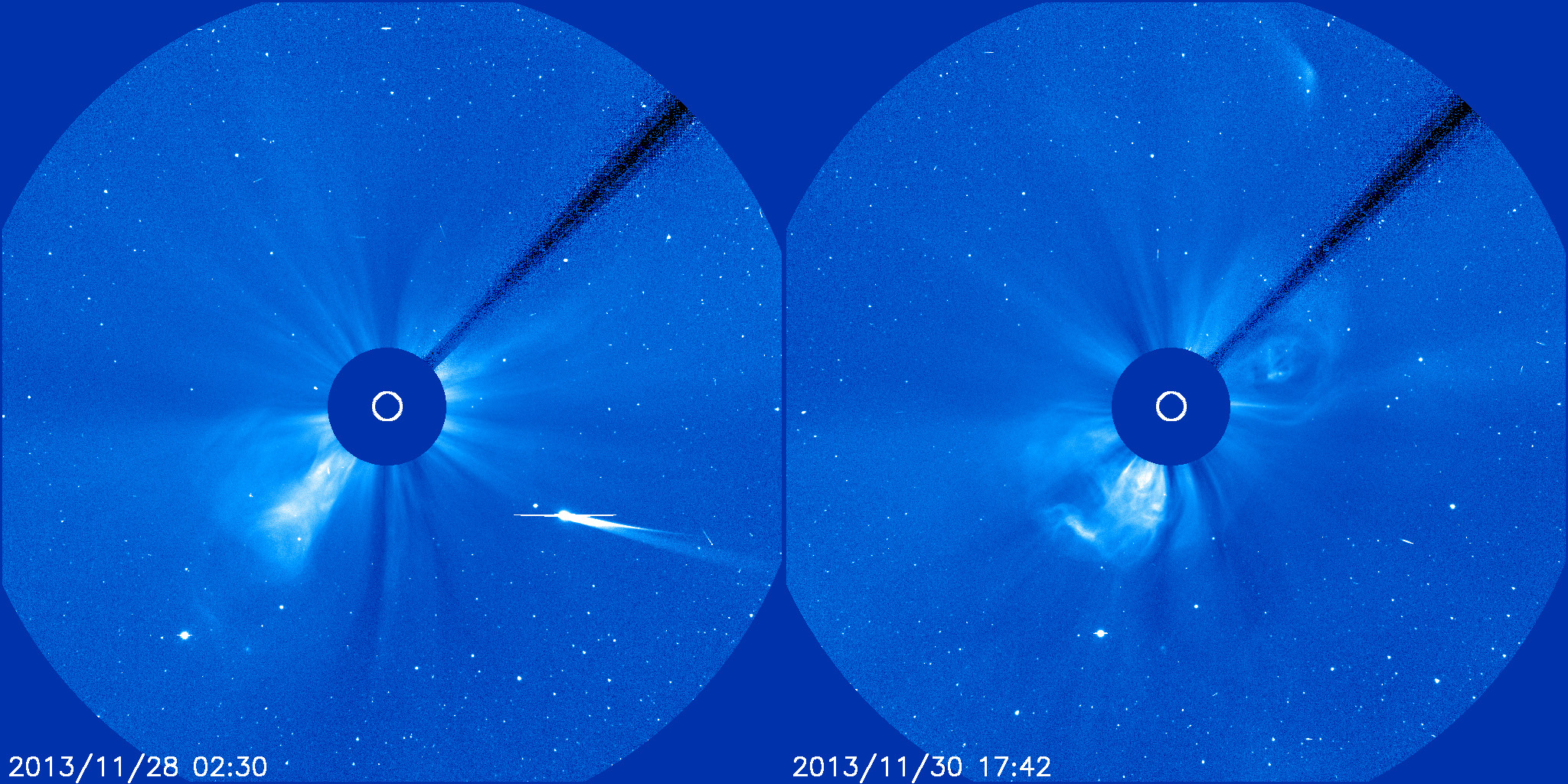}
 	   	 	 	\caption{\small {Images recorded by LASCO C3 camera. In the left image it is possible to observe the passage of the sungrazer comet ISON at perihelion on 28 November 2013 (Taken from \cite{IMAGES}).}}
 	   	 	 	\label{fig:ison}
 	   	 	 	\end{figure}

SOHO was launched on the 2nd of December 1995 becoming operative in 1996 in the Lagrangian Point L1 of Earth's orbit. This mission still provides images on a daily basis of the solar corona in 2014.

	\begin{figure}[h]
 	   	 	 	\centering
 	   	 	 	\includegraphics[width=0.95\linewidth]{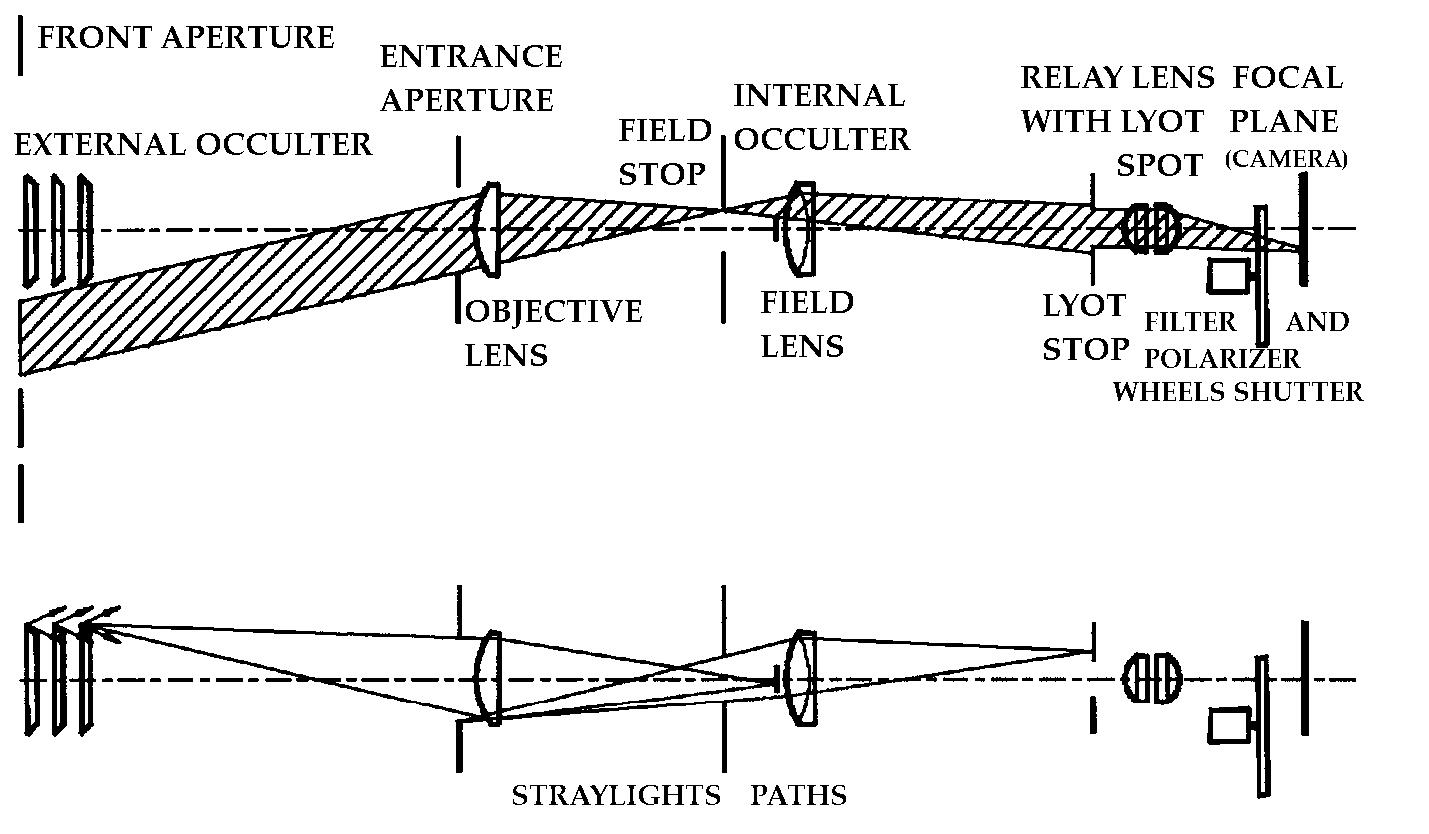}
 	   	 	 	\caption{\small {Optical scheme of C3 coronograph. Top diagram: the image paths. Bottom diagram: raytracing for the suppression of straylights. Adapted from \cite{LASCO_homepage}.}}
 	   	 	 	\label{fig:optical_scheme}
 	   	 	 	\end{figure}

\section{Data}
\subsection{Data Collecting}
\begin{figure}[H]
 	   	 	 	\centering
 	   	 	 	\includegraphics[width=0.95\linewidth]{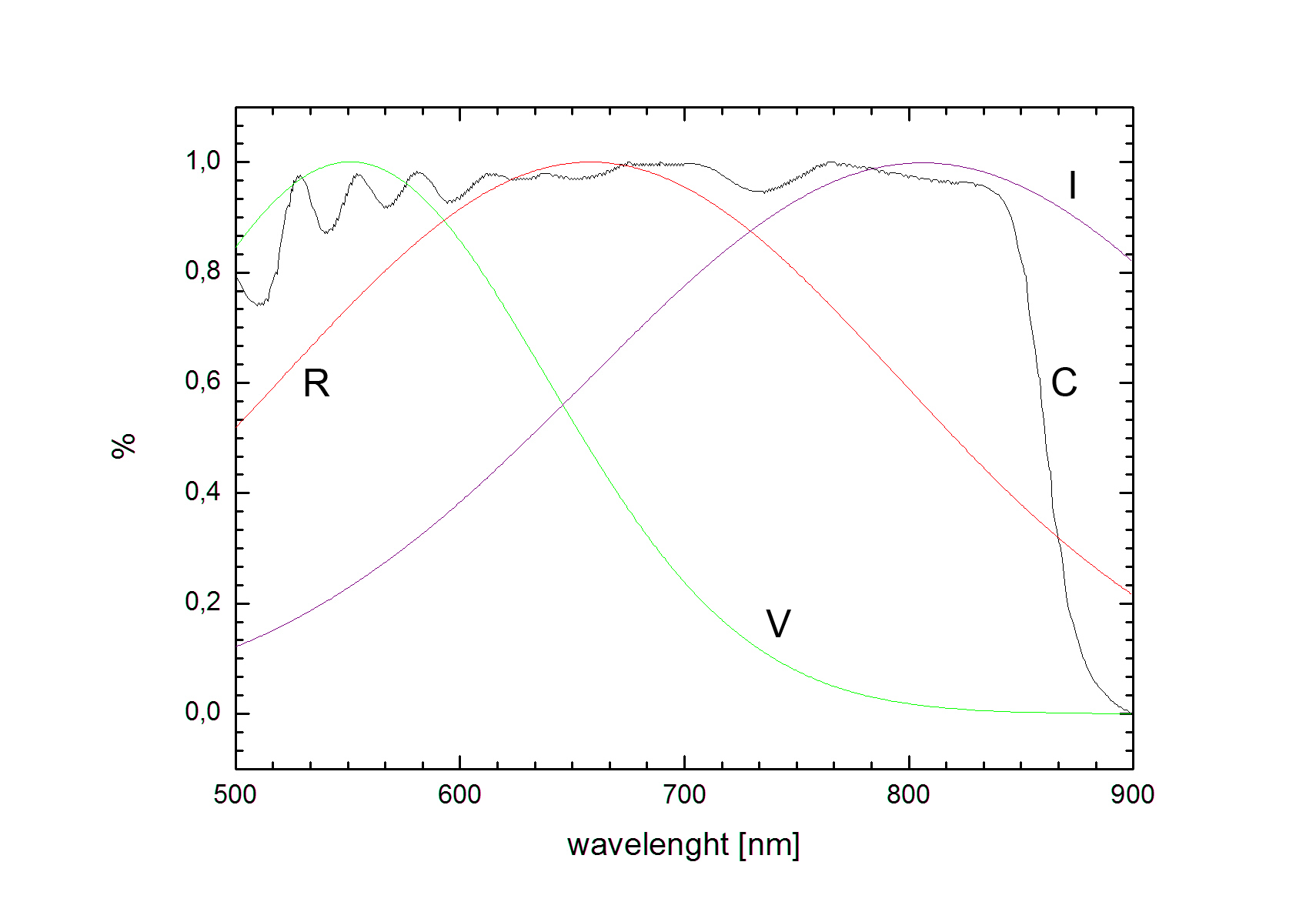}
 	   	 	 	\caption{\small {C: Clear band Quantum Efficiency of SOHO LASCO/C3; V: Johnson V-band; R: Johnson R-band; I: Johnson I-band.}}
 	   	 	 	\label{fig:clear_band}
 	   	 	 	\end{figure}

Delta Scorpii is in the field of view of SOHO for approximately 6 days around 30th of November. We analyzed the FITS images available from November 1996 to December 2013 from LASCO C3 camera archive, namely LASCO/EIT \cite{LASCO_EIT}; the images are in the \emph{Clear band} (see \ref{fig:clear_band}).

Being Delta Scorpii a blue giant, we focused on the V magnitude. The quantum efficiency of the LASCO C3 camera (\ref{fig:clear_band}) covers a wider range of wavelenghts than the Johnson V-band; the conversion between Johnson V magnitude and LASCO C3 one has been numercally evaluated as $M_{V}= M_{C3}+0.093$ for a 27000 K black body\cite{temp}.

FITS are not elaborated images, without loss of information due to image compression.
Approximately 2400 images have been analyzed for making differential photometry, using Beta Scorpii and Pi Scorpii as reference stars because their color index B-V is similar to the one of Delta Scorpii, 
avoiding data within the solar Corona because of the poorer contrast.

	\begin{figure}[H]
 	   	 	 	\centering
 	   	 	 	\includegraphics[width=0.85\linewidth]{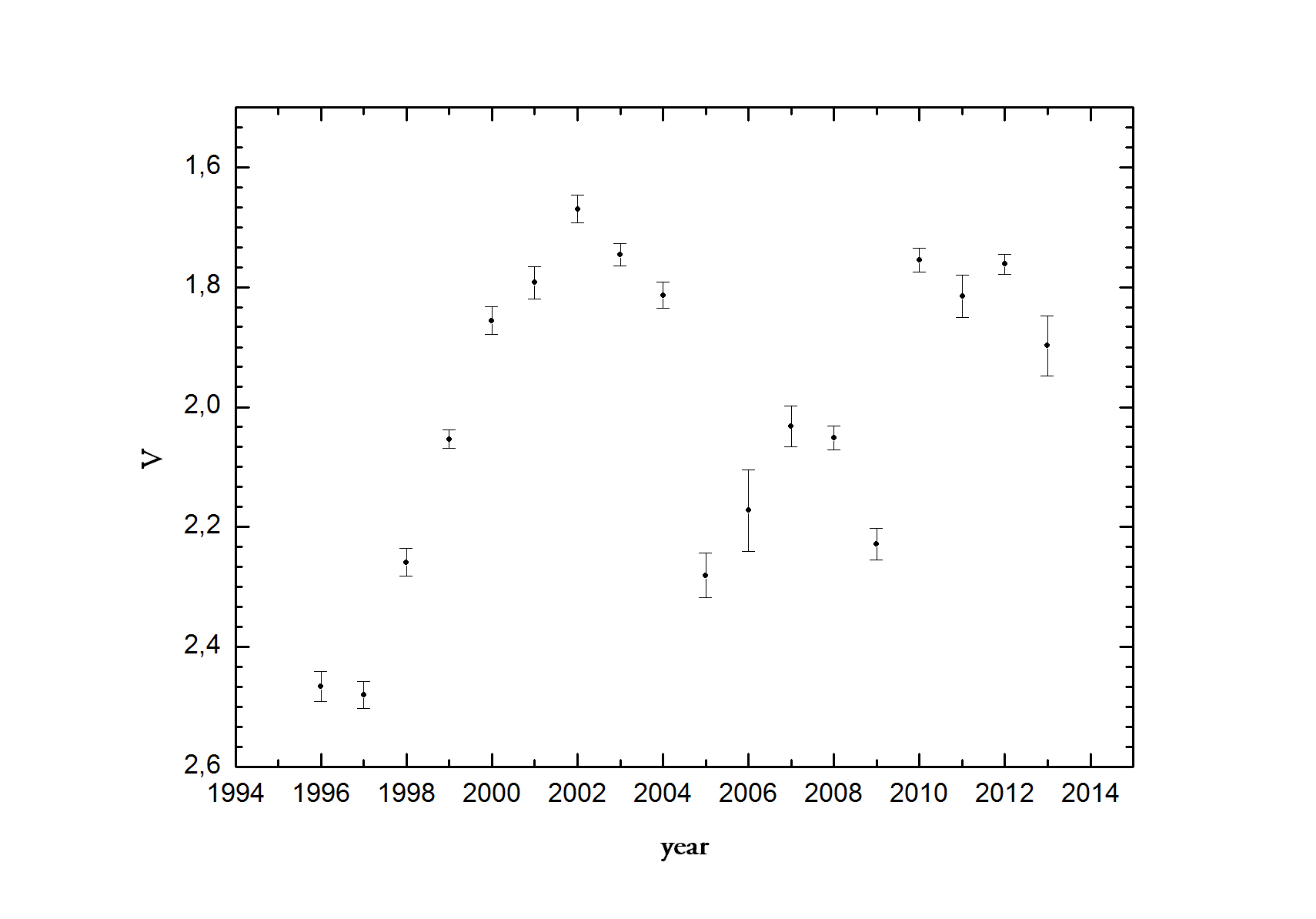}
 	   	 	 	\caption{\small {Magnitudes in Johnson V-band vs year. Each point is the mean of the magnitudes aquired from the 28th of November to the 1st of December of each year with its standard deviation.}}
 	   	 	 	\label{fig:fig1}
 	   	 	 	\end{figure}

\begin{figure}[H]
	\centering
	\includegraphics[width=0.85\linewidth]{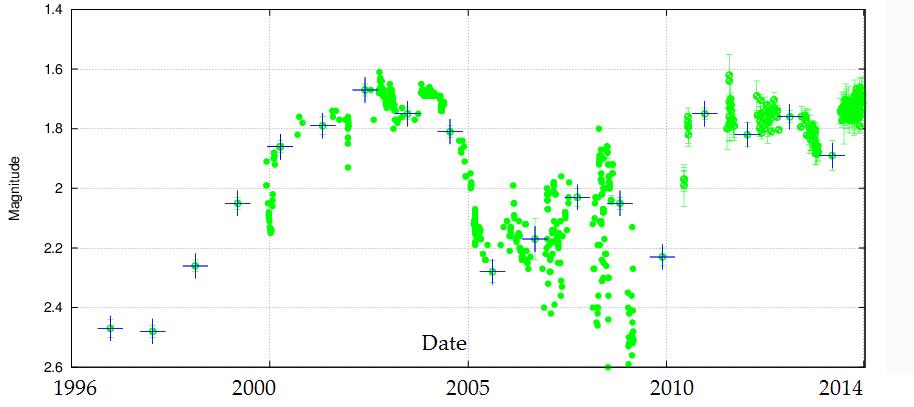}
	\caption{\small {Comparison of SOHO LASCO C3 data on Delta Scorpii with the AAVSO data in the V-band. Green dots: AAVSO data. Blue crosses:
			C3 data.}}
	\label{fig:confronto}
\end{figure}

\subsection{Dispersion in magnitude, vignetting and diffraction in C3}

Between the 28th of November and the 1st of December, the data show a typical 0.2 magnitudes dispersion.
A plot shows eventually the presence of significant trends. Their absence validates the use of the average value.

	\begin{figure}[h]
 	   	 	 	\centering
 	   	 	 	\includegraphics[width=0.95\linewidth]{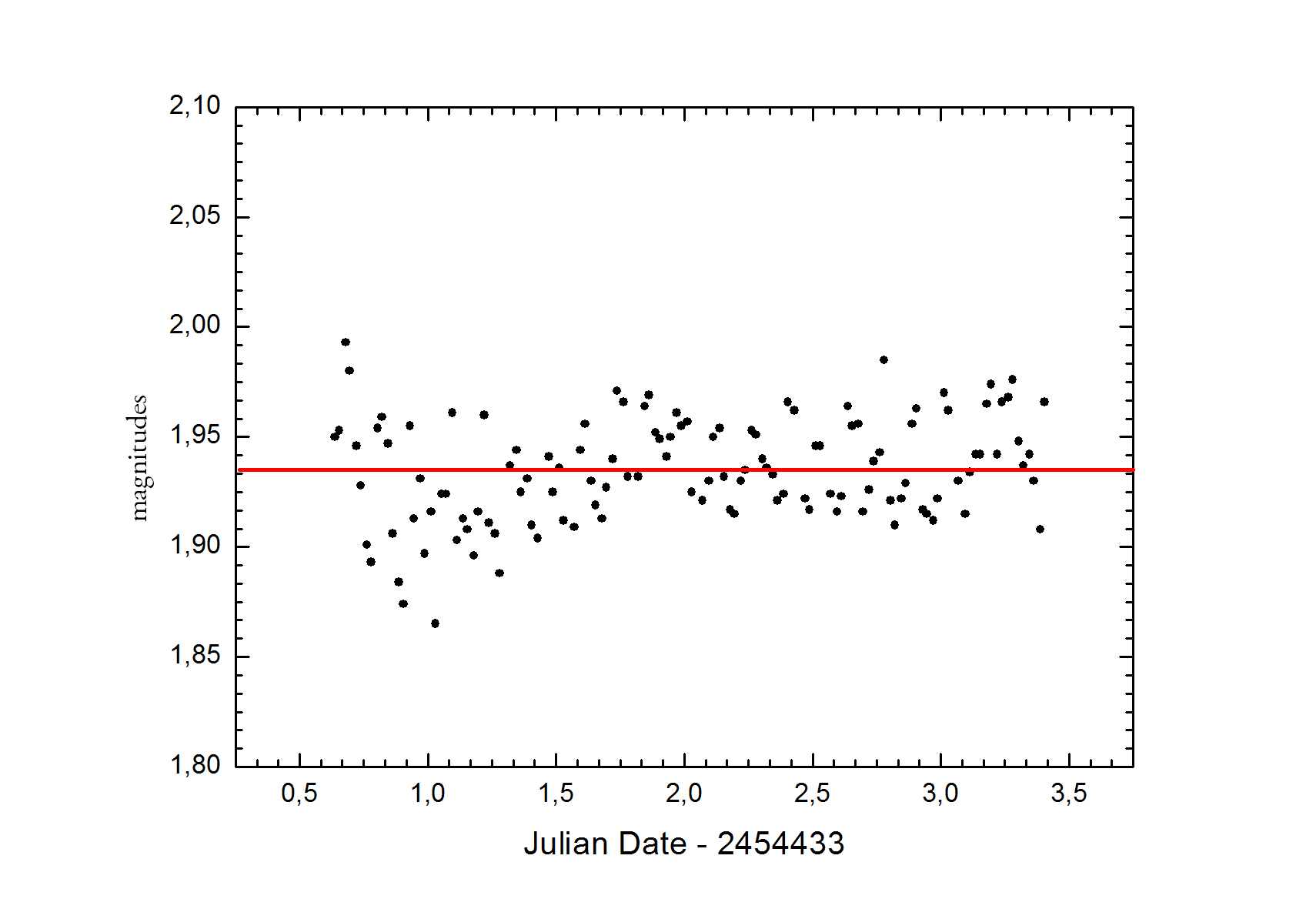}
 	   	 	 	\caption{\small {V magnitude of Delta Scorpii measured during four days. The line represents the arithmetical mean. Data from 28 Nov 2007.}}
 	   	 	 	\label{fig:dati}
 	   	 	 	\end{figure}

Vignetting and diffraction caused by the coronograph are responsible for the magnitude dispersion.

\begin{figure}[H]
 	   	 	 	\centering
 	   	 	 	\includegraphics[width=0.8\linewidth]{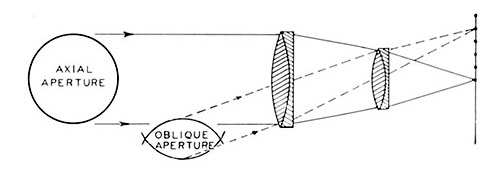}
 	   	 	 	\caption{\small {The vignetting of oblique beams of light by a lens. As not all the oblique rays reach the focal plane, we have a progressive darkening of the image at the edges \cite{KINGSLAKE}.}}
 	   	 	 	\label{fig:vignetting}
 	   	 	 	\end{figure}

The presence of a stop in an optical system causes the vignetting effect, namely the progressive darkening on the edge of an image (See Fig. \ref{fig:vignetting}).

\begin{figure}[H]
 	   	 	 	\centering
 	   	 	 	\includegraphics[width=0.95\linewidth]{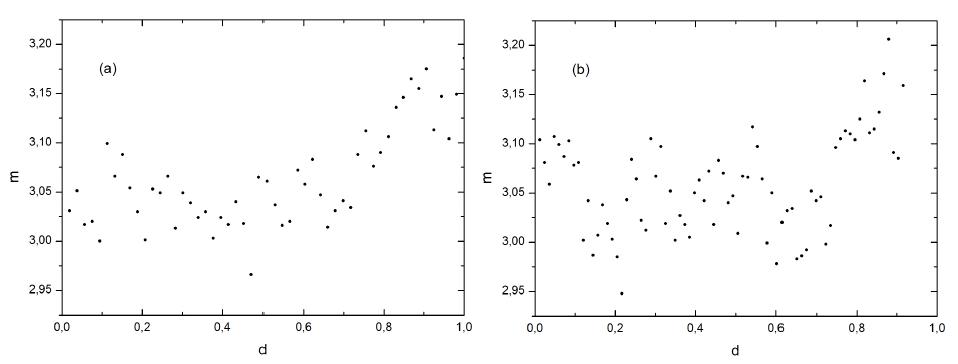}
 	   	 	 	\caption{\small {Effect of vignetting on the images: Pi Scorpii magnitude vs its normalized radial distance; the edge is at d=1. (a) Data relative to year 2004. (b) Data relative to year 2005. Different years are in order to exclude possible spurious random effects.}}
 	   	 	 	\label{fig:fig2}
 	   	 	 	\end{figure}

To measure this effect, we studied the motion of Pi Scorpii across the image (always centered on the Sun) towards the edge of the field of view, measuring its magnitude at every step. In Fig. \ref{fig:fig2} it is plotted the magnitude of Pi Scorpii in function of its distance from the edge.

From the Fig. \ref{fig:fig2} we see that vignetting produces a 0.15 magnitudes scatter in our data. 

Since the coronograph includes various obstacles and stops between the detector and the light source, they cause appreciable diffraction patterns according to the Babinet's principle. 

\begin{figure}[H]
 	   	  	   	 	 	\centering
 	   	  	   	 	 	\includegraphics[width=0.4\linewidth]{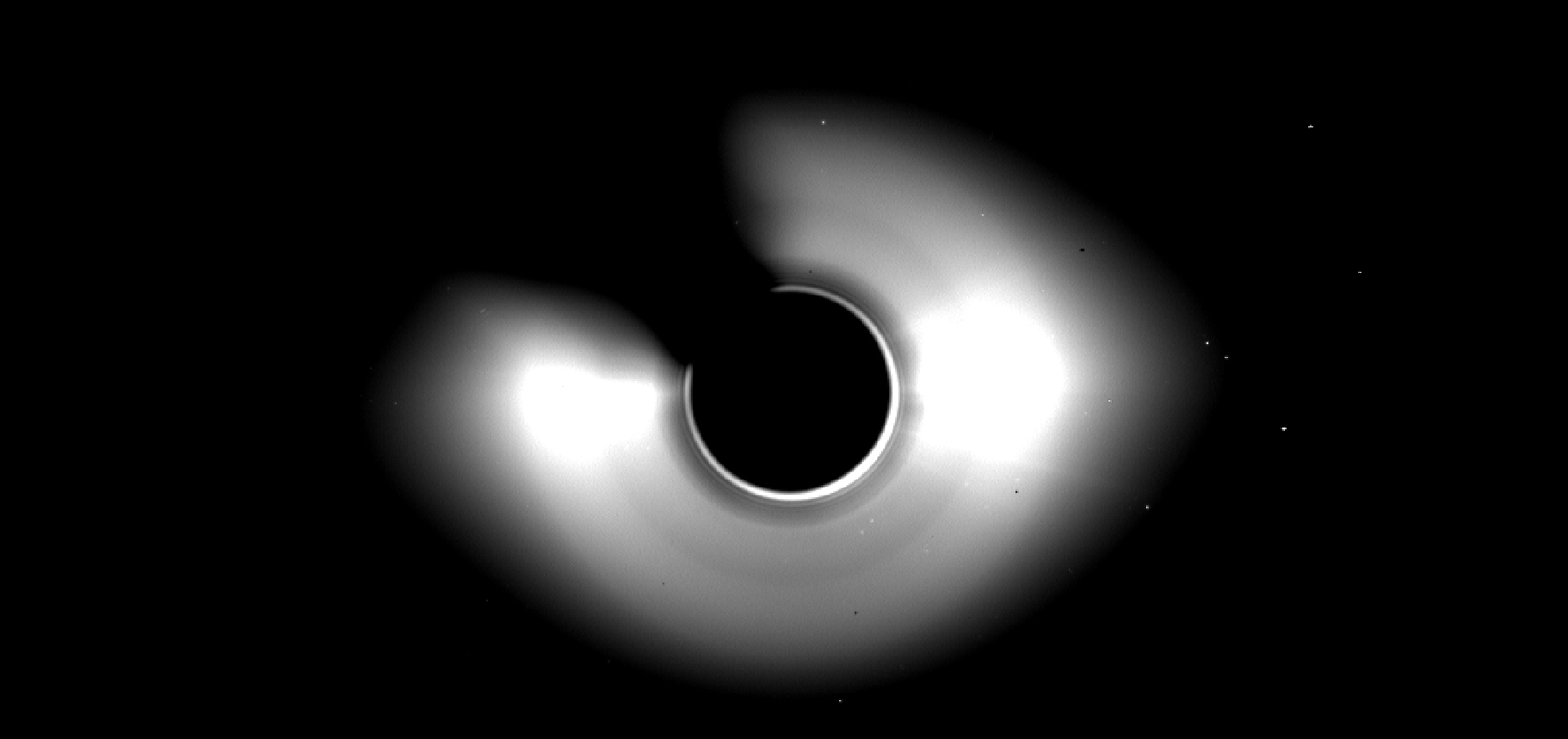}
 	   	  	   	 	 	\qquad\qquad
 	   	  	   	 	 	\includegraphics[width=0.4\linewidth]{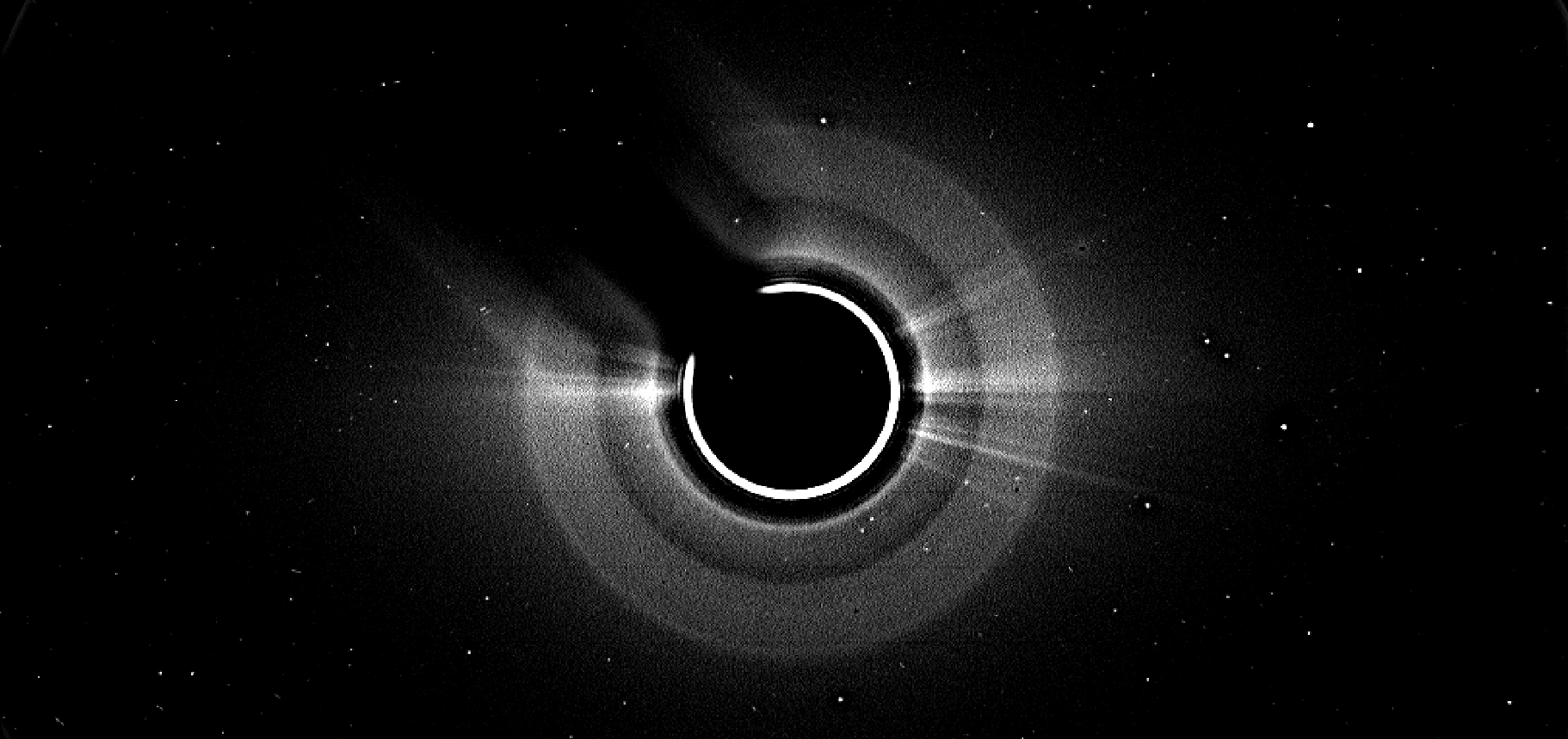}
 	   	  	   	 	 	\caption{\small {Left panel: FITS image not elaborated. Right panel: The same image after enhancing the contrast.}}
 	   	  	   	 	 	\label{fig:immagini}
 	   	  	   	 	 	\end{figure}

In a generic image in the left panel of Fig. \ref{fig:immagini}, after enhancing the contrast as in the right panel of Fig. \ref{fig:immagini}, we traced various intensity profiles, all passing from the centre of the image, with different azimuth. 
The mean of these profiles is a clean representation of this diffraction pattern, obtained by using the Sun as a source. 
  	 
\begin{figure}[H]
 	 	   	 	 	\centering
 	 	   	 	 	
 	 	   	 	 	\includegraphics[width=0.95\linewidth]{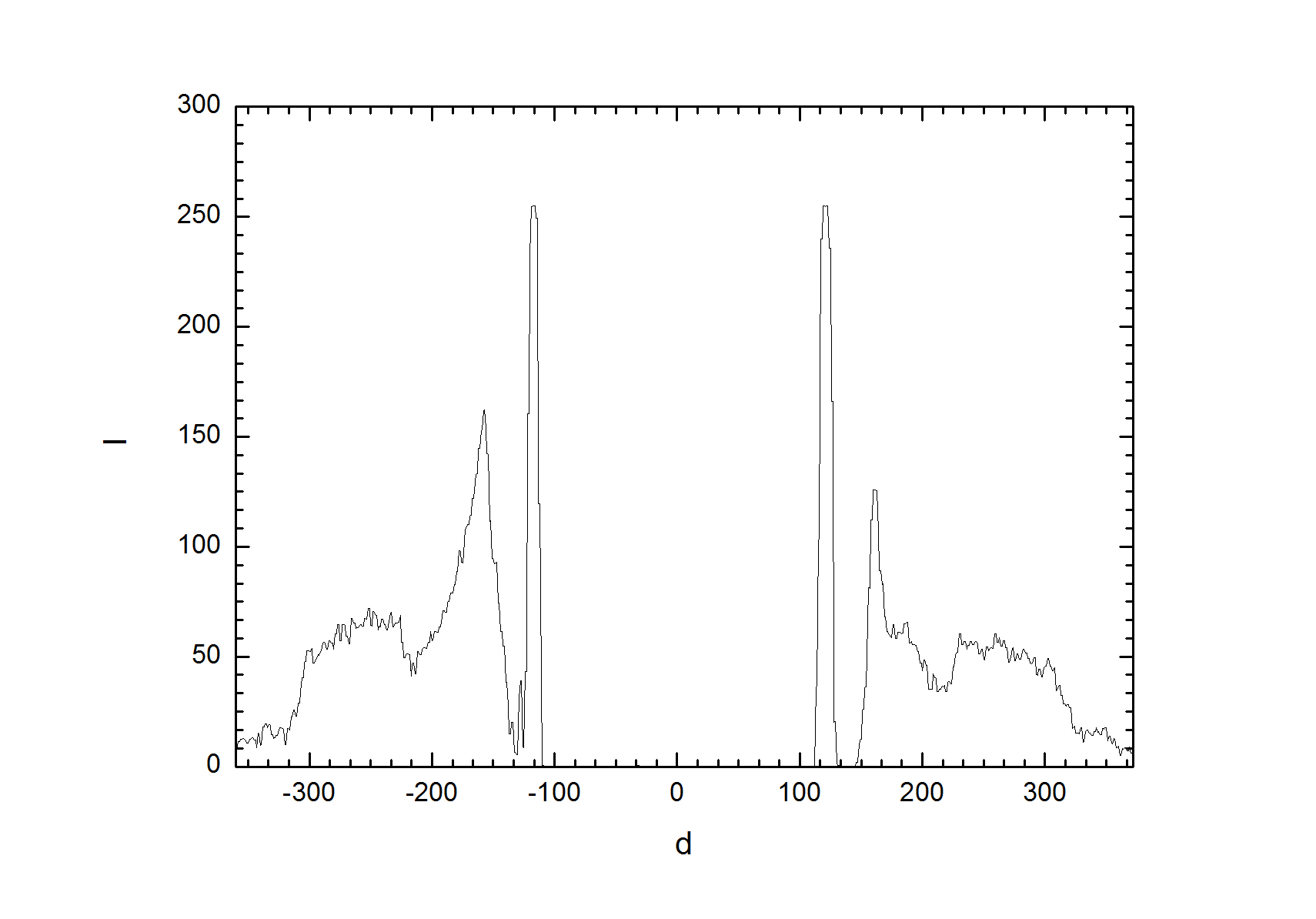}
 	 	   	 	 	\caption{\small {Diffraction profile due to the coronograph. In the graph it is plotted the intensity I (from 0 to 255) vs radial coordinate d from the center.}}
 	 	   	 	 	\label{fig:fig3}
 	 	   	 	 	\end{figure}

The average azimuthal diffraction pattern is shown in Fig.\ref{fig:fig3}.
 	   	  	 	 	 	   	 	 	 	   	  	 	
Measuring the magnitude of a star in two different regions: $m$ where the effect of diffraction is negligible and $m'$ where the effect is strong we evaluated the difference $\delta m=m'-m\simeq 0.154$mag. 	 	   	 	 	

The contribution of both vignetting and diffraction accounts for the spread of our experimental data. 

\subsection{Conclusions}

We analyzed Delta Scorpii data from SOHO from November 1996 to December 2013, using FITS images of LASCO C3 camera. 
As reference stars we used Beta and Pi Scorpii because of their color index B-V and their closeness to Delta Scorpii. We discarded the images when the stars were too close to the solar corona and to the regions with maximum vignetting and diffraction. Their combined effect to the magnitude estimates of the star produce a scatter within 0.2 magnitudes. 

The extension back to 1996 of Delta Scorpii allows to know $25\%$ more of its intriguing light curve thanks to these reliable satellite data.
The minimum of 1996-1997 at 2.5 magnitude observed with SOHO enforces the 11 year periodicity related to the orbital period of the companion\cite{SIGISMONDI}.

Using SOHO satellite to study other zodiacal stars, during their period of conjunction with the Sun, may open new perspectives
in the domain of stellar variability, filling the yearly gap of usual unobservability of the stars.

{\bf{Acknowledgments:}} 
Costantino Sigismondi is grateful to Alexandre Amorim, astronomer of Florianopolis, Brazil. 
He drove my attention to the comet ISON during the IAU/LARIM 2013 meeting, showing the image of SOHO with delta Scorpii. Our observations of the comet ISON during its grazing perihelion of Nov 28, 2013 documented in Fig.\ref{fig:ison} are described in the number of December 2013 of the bulletin Observe!\cite{Observe!}

%\pagebreak


\begin{thebibliography}{Bibliography}
\bibitem{LASCO_EIT}LASCO/EIT Images Query Form\\
http://sharpp.nrl.navy.mil/cgi-bin/swdbi/lasco/img-short/form.
\bibitem{LASCO_homepage}Homepage LASCO\\
http://lasco-www.nrl.navy.mil.
\bibitem{SOHO_homepage}Homepage SOHO\\
http://sohowww.nascom.nasa.gov.
\bibitem{OTERO}S. Otero\\
http://www.aavso.org/vsots\_delsco.
\bibitem{SIGISMONDI}C. Sigismondi, \emph{Delta Scorpii 2011 periastron: worldwide observational campaign and preliminary photometric analysis}, arXiv:1107.1107v1.
\bibitem{BRUECKNER}G. E. Brueckner et al., \emph{The large angle spectroscopic coronograph (LASCO)}, Solar Physics, Dec 1995, Volume 162, Issue 1-2, pp 357-402.
\bibitem{IMAGES}SOHO data retrieval\\
http://sohodata.nascom.nasa.gov/cgi-bin/data\_query.
\bibitem{Observe!}Observe! Bulletin\\
{\rm http://www.geocities.ws/costeira1/neoa/observe.htm}
\bibitem{KINGSLAKE}R. Kingslake, \emph{Lenses in Photography}, Garden City Books, New York, 1951; Second Edition, A. S. Barnes, New York, 1963.
\bibitem{GCVS}General Catalogue of Variable Stars\\
http://www.sai.msu.su/gcvs/gcvs/index.htm.
\bibitem{VARSAO}http://varsao.com.ar/delta\_Sco.htm.
\bibitem{MINDSPRING}http://www.mindspring.com/$\sim$feez/Star.htm.
\bibitem {temp} D. P. K. Banerjee, P. Janardhan, and N. M. Ashok, \emph{Near infra-red and optical spectroscopy of Delta Scorpii}, India, Physical Research Laboratory, Navrangpura, 2001.
\bibitem{aavso}AAVSO web site\\ http://www.aavso.org.


\end{thebibliography}
\end{document}